\documentclass[twocolumn,showpacs,preprintnumbers,amsmath,amssymb,%
prl,a4paper,floatfix,nofootinbib]{revtex4-1}  

\usepackage{graphics}
\usepackage{graphicx}
\usepackage{epsfig}
\usepackage{color}
\usepackage{longtable}
\usepackage{hyperref}
\usepackage{latexsym}
\usepackage{amsmath}
\usepackage{amsthm}
\usepackage{amsfonts}
\usepackage{amssymb}
\usepackage{dsfont}
\usepackage{bm}
\usepackage{graphics,psfrag}
\usepackage{graphicx,psfrag}
\usepackage[nice]{nicefrac}

\newcommand{\ot}{\ensuremath{\frac{1}{2}}}
\newlength{\mylenC}
\begin{document}
%
\title{Low-lying $\Lambda$ Baryons from the Lattice}
\author{Georg P.~Engel$^1$, 
C.~B.~Lang$^1$, 
and 
Andreas Sch\"afer$^2$\\
\vspace*{2mm}
(BGR [Bern-Graz-Regensburg] Collaboration)
\vspace*{2mm}\\}

\affiliation{
$^1$Institut f\"ur Physik, FB Theoretische Physik, Universit\"at
Graz, A--8010 Graz, Austria\\
$^2$Institut f\"ur Theoretische Physik, Universit\"at
Regensburg, D--93040 Regensburg, Germany
}

\date{\today}

\begin{abstract}
In a lattice QCD calculation with two light dynamical Chirally Improved (CI) quarks
we determine ground state and some excited state masses in all four $\Lambda$ baryon
channels 1/2$^\pm$ and 3/2$^\pm$. We perform an infinite volume extrapolation and
confirm the widely discussed $\Lambda(1405)$. We also analyze the amount of
octet-singlet mixing, which is helpful in comparing states with the quark model.
\end{abstract}

\pacs{11.15.Ha, 12.38.Gc}
\keywords{Hadron spectroscopy, dynamical fermions}

\maketitle

One of the central aims of hadron spectroscopy is to understand the
spin-flavor-parity structure of the excitation spectra for different  quantum
numbers. It seems in particular that the nucleon and $\Lambda$  spectra show
significant differences which, if properly understood, should  illuminate the
influence of quark mass and flavor on hadron structure. The lowest $\Lambda(\ot^-)$ mass
lies below the Roper-like $\Lambda(1600,\ot^+)$
and the negative parity nucleon $N^*(1535)$;  unlike the nucleon sector it
does have standard level ordering lying between  the positive parity ground state
and the first positive parity excitation.  $\Lambda$ baryons can be flavor singlets
or octets or, due to the difference in  light and strange quark masses, mixtures of
both. Various continuum model studies discuss that mixing. 
This is the first lattice
QCD analysis of the $\Lambda$ baryons for dynamical  quarks, that includes {\em all
four}, namely the $J^P=\ot^\pm$- and  $\frac{3}{2}^\pm$-channels. We obtain ground
states compatible with experiment in three of those.  We also study the infinite
volume limit and give for the first time  lattice results on singlet-octet
composition for all four sectors,  obtaining the mass of the $\Lambda(1405)$
and confirming its flavor singlet nature.

Lattice studies in the quenched case generally had problems to find the low lying
$\Lambda(1405)$. Even a study with two dynamical light quarks
\cite{Takahashi:2009ik,Takahashi:2009bu} found too large mass values. Only recent
results \cite{Menadue:2011pd} obtained with  PACS-CS $(2+1)$-flavor dynamical quarks
configurations \cite{Aoki:2008sm} show a level ordering which is compatible with experiment.

We study the baryons for a set of seven ensembles with two dynamical Chiral Improved
(CI) quarks \cite{Gattringer:2000js,Gattringer:2000qu}. 
The CI
fermion action consists of several hundred terms and obeys the Ginsparg-Wilson
relation in a truncated approximation. The pion mass ranges from 255 to 596 MeV, the
lattice spacing lies between 0.1324 and 0.1398 fm. The bulk of our results were
obtained  for lattices of size $16^3\times 32$. For two ensembles with light
pion masses (255 and 330 MeV) we also used $24\times 48$  lattices. Thus $m_\pi L$
which controls the finite size effects is 4.08 and 5.61 in these two situations.
Further details on the action and our simulation, as well as results for mesons, 
can be found in \cite{Engel:2010my,Engel:2011aa}.
The strange quark is
introduced as a valence quark and its mass fixed by the $\Omega$-mass. 

The Dirac and flavor structure of the interpolating fields used in our
study is motivated by the quark model 
\cite{Isgur:1978xj,Isgur:1978wd}, see also \cite{Glozman:1995fu}. Within the
relativistic quark model there have been many determinations of the
hadron spectrum, based on confining potentials and different assumptions
on the hyperfine interaction (see, e.g., \cite{Capstick:1986bm,Glozman:1997ag,Loring:2001kx}).
The singlet, octet and decuplet attribution \cite{Glozman:1995fu} of the states
has been evaluated based on such model calculations, e.g., in \cite {Melde:2008yr}
(see also the summary in \cite{Nakamura:2010zzi}).

For the $\Lambda$ baryons we use sets of up to 24 interpolating fields in each
quantum channel. The singlet and octet combinations of Table
\ref{tab:baryon:interpol:1} are combined with three possible choices of Dirac
matrices $(\Gamma_1, \Gamma_2)=$ $(\mathds{1}, C\gamma_5)$, $(\gamma_5, C)$     and 
$(\mathrm{i} \mathds{1},C\gamma_t\gamma_5)$ (denoted by $\chi_1$, $\chi_2$ and
$\chi_3$ for short) for $J=\ot$ and eight combinations of Gaussian smeared quarks
\cite{Gusken:1989ad,Best:1997qp} with two smearing widths $(n,w)$. The operator
numbering is given in Table \ref{tab:baryon:interpol:3}. All interpolators are
projected to definite parity and all  Rarita-Schwinger fields (spin $\frac{3}{2}$
interpolators in Table \ref{tab:baryon:interpol:3}) are projected to definite spin
$\frac{3}{2}$ using the continuum formulation of the Rarita-Schwinger projector
\cite{Lurie:1968}. $C$ denotes the charge conjugation matrix, $\gamma_t$ and
$\gamma_i$ the time and the spatial direction Dirac matrices.

For point like quark fields, Fierz identities reduce the actual number 
of independent operators (see, e.g., \cite{Chen:2008qv}). 
In particular, there are no non-vanishing point-like interpolators for $\Delta(\ot)$
and singlet $\Lambda(\frac{3}{2})$.  We use different quark smearing widths such that
the Fierz identities do not apply.
Hence $\chi_1$, $\chi_2$ and  $\chi_3$ are
all independent, for $J=\frac{3}{2}$ all interpolators are non-vanishing.

From the cross-correlation matrix $C_{ik}(t)=\langle  O_i(t) O_k(0)^\dagger\rangle$
we obtain the energy levels with help of the variational method
\cite{Luscher:1990ck,Michael:1985ne}. One  solves the generalized eigenvalue problem
$C(t) \vec u_n(t)=\lambda_n(t) C(t_0) \vec u_n(t) $ in order to approximately
recover the energy eigenstates $|n\rangle$. The eigenvalues
$\lambda_n(t)\sim\exp(-E_n\,t)$ allow us to get the energy values  and the
eigenvectors serve as  fingerprints of the states, indicating their content in terms
of the lattice interpolators. The quality of the results depends on the statistics
and the provided set of lattice operators. The dependence on $t_0$ is used to study
the systematic error; in the final analysis we use $t_0=1$ (with the origin at 0).
The statistical error is determined with single-elimination jack-knife. For the fits
we use single exponential behavior but check the stability with double exponential
fits; we take the correlation matrix for the correlated fits from the complete
sample \cite{Engel:2011aa}. 

For the extrapolation to the physical pion
mass we fit to the leading order chiral behavior, which is linear in $m_{\pi}^2$.
Two ensembles (at pion masses 255 MeV and 588 MeV) suffer from a slight mistuning of the strange quark mass, which are therefore discarded 
in the extrapolation to the physical pion mass, whenever the effects are found significant.
This is the case for the lowest energy levels in each channel (three lowest ones in $\ot^{-}$). 
The quoted systematic errors for these levels include the corresponding deviation and the dependence of the energy levels on the choice of interpolators and fit ranges for the eigenvalues.  

In the present study we are restricted to 3-quark operators for the baryon.
Note that ideally one should take into account also meson-baryon interpolators (see, e.g., the discussion
in \cite{Oller:2000fj}). This leads to many more contributing graphs and 
necessitates also the inclusion of backtracking quark loops.
The related computational and algorithmic effort prevented such
lattice calculations so far, although such studies are in progress \cite{Mohler:2012nh}.
Due to sea quarks, in principle, 3-quark operators have overlap with meson-baryon states as well.
The corresponding coupling was however found to be weak in actual simulations \cite{Engel:2010my,Bulava:2010yg}.
We will argue below that in particular in the s-wave channels we find hints of such coupling even for our interpolators. 

\begin{table}[t]
\begin{center}
\begin{tabular}{cccc}
\hline
\hline
Spin		& Flavor  	& Name				& Interpolator \\
\hline
$\frac{1}{2}$	& Singlet	& $\Lambda_{1/2}^{(1,i)}$	& $\epsilon_{abc} \Gamma^{(i)}_1 u_a ( d_b^T \Gamma^{(i)}_2 s_c - s_b^T \Gamma^{(i)}_2 d_c) 				$ \\ 
		&			& 				& $\,+ \, \mbox{cyclic permutations of}\;  u, d, s  									$ \\
$\frac{1}{2}$	& Octet		& $\Lambda_{1/2}^{(8,i)}$	& $\epsilon_{abc} \Big[ \Gamma^{(i)}_1 s_a ( u_b^T  \Gamma^{(i)}_2 d_c - d_b^T  \Gamma^{(i)}_2 u_c ) 			$ \\
		&			& 				& $\, + \; \Gamma^{(i)}_1 u_a ( s_b^T  \Gamma^{(i)}_2 d_c) - \Gamma^{(i)}_1 d_a ( s_b^T  \Gamma^{(i)}_2 u_c) \Big]	$ \\
$\frac{3}{2}$	& Singlet	& $\Lambda_{3/2}^{(1,i)}$	& $\epsilon_{abc} \gamma_5 u_a ( d_b^T C \gamma_5 \gamma_i s_c - s_b^T C \gamma_5 \gamma_i d_c) 			$ \\ 
		&			& 				& $\,+ \, \mbox{cyclic permutations of}\;  u, d, s  									$ \\
$\frac{3}{2}$	& Octet		& $\Lambda_{3/2}^{(8,i)}$	& $\epsilon_{abc} \Big[ \gamma_5 s_a ( u_b^T C \gamma_5 \gamma_i d_c - d_b^T  C \gamma_5 \gamma_i u_c ) 		$ \\
		&			& 				& $\, + \; \gamma_5 u_a ( s_b^T  C \gamma_5 \gamma_i d_c) - \gamma_5 d_a ( s_b^T  C \gamma_5 \gamma_i u_c\Big]		$ \\
\hline
\hline
\end{tabular}
\end{center}
\caption[Baryon interpolators:]{
Baryon interpolators: 
The possible choices for the Dirac matrices $\Gamma_{1,2}^{(i)}$ in the spin \ot channels are discussed in the text. Summation convention applies; for spin $\frac{3}{2}$ observables, the open Lorentz index (after spin projection) is summed after taking the expectation value of correlation functions.
}
\label{tab:baryon:interpol:1}
\end{table}

\begin{table}[t]
\begin{center}
\begin{tabular}{ccccc}
\hline
\hline
Quark		& \multicolumn{4}{c}{Numbering of associated interpolators} \\
smearing types	& ~~~~~$\Lambda_{3/2}^{(1,i)}$~					&~$\Lambda_{3/2}^{(8,i)}$~			& ~$\Lambda_{1/2}^{(1,i)}$~			& $\Lambda_{1/2}^{(8,i)}$	\\
\hline
(nn)n		& ~~~1								& ~9					& 1,9,17				& 25,33,41	\\
(nn)w		& ~~~2								& 10					& 2,10,18				& 26,34,42	\\
(nw)n		& ~~~3								& 11					& 3,11,19				& 27,35,43	\\
(nw)w		& ~~~4								& 12					& 4,12,20				& 28,36,44	\\
(wn)n		& ~~~5								& 13					& 5,13,21				& 29,37,45	\\
(wn)w		& ~~~6								& 14					& 6,14,22				& 30,38,46	\\
(ww)n		& ~~~7								& 15					& 7,15,23				& 31,39,47	\\
(ww)w		& ~~~8								& 16					& 8,16,24				& 32,40,48	\\
\hline
\hline
\end{tabular}
\end{center}
\caption[Baryon interpolators: Quark smearing types]{ Baryon interpolators:
Quark smearing types (n/w for narrow/wide) and naming convention for the interpolators in the
different channels. The three columns for the $J=\ot$ interpolators refer to $\chi_1$--$\chi_3$.}
\label{tab:baryon:interpol:3}
\end{table}

\begin{figure}
\noindent\includegraphics[width=\mylenC,clip]{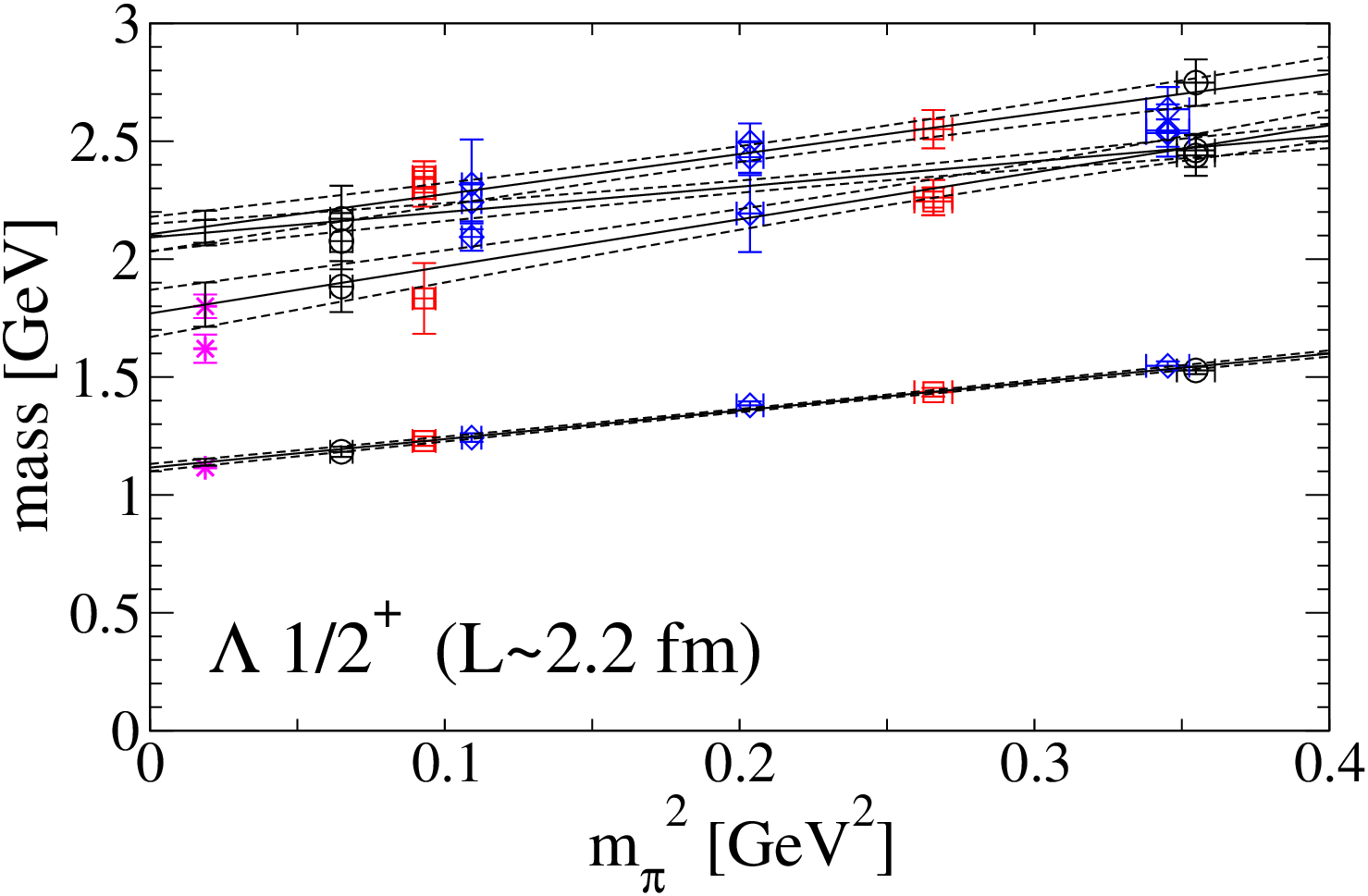}
\caption{
Energy levels for $\Lambda$ spin $\ot^+$ in a finite box of linear size
$L\approx 2.2$ fm. Stars denote the experimental values
\cite{Nakamura:2010zzi}, other symbols denote results from the simulation. The
full lines show the extrapolations linear in the $m_\pi^2$, the broken
curves indicate the statistical error bands.}
\label{fig:lambda_1half_pospar}
\end{figure}

\begin{figure}
\includegraphics[width=\columnwidth,clip]{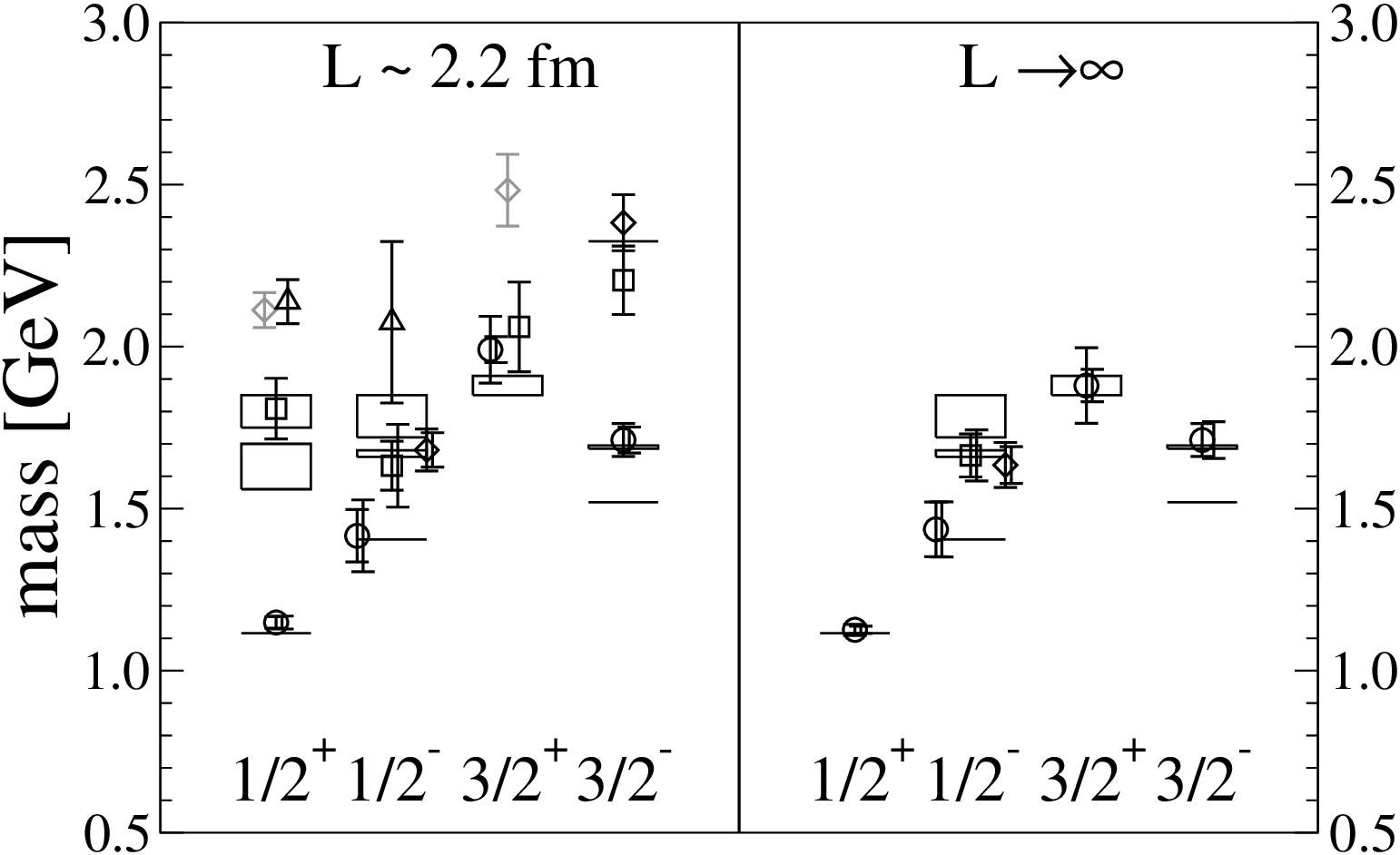}
\caption{
Energy levels extrapolated to the physical pion mass
in finite volume $L\approx 2.2$ fm (lhs) and low lying levels after
infinite volume extrapolation (rhs). The horizontal lines or boxes represent
experimentally known states \cite{Nakamura:2010zzi}, where the box size
indicates the statistical uncertainty of 1$\sigma$.  The statistical uncertainty
of our results is indicated by error bars of 1$\sigma$, second bars closely to the
right indicate systematic errors estimated from the use of different sets of
interpolators, fit ranges of the eigenvalues and the strange quark mass. Grey symbols denote a poor
$\chi^2$/d.o.f. $>3$ of the chiral fits.
}
\label{fig:summary}
\end{figure}

\paragraph{$J^P=\ot^{+}$, Finite volume:}
In  Fig.~\ref{fig:lambda_1half_pospar} we show results for the four lowest
eigenstates for interpolator set (1,2,11,20,25,26,33,34,43).  After
extrapolation to the physical point our lowest energy level agrees well with the
experimental $\Lambda(1116)$. The systematic error estimated from different
combinations of interpolators and fit ranges is indicated in the summary
Fig.~\ref{fig:summary}. Analyzing
the eigenvectors, we find that the ground state is dominated by octet interpolators
of Dirac structure $\chi_1$ and $\chi_3$ (in agreement with a relativistic
quark model calculation \cite{Melde:2008yr}).
Our first excitation is dominated by singlet interpolators (first Dirac structure)
matching the $\Lambda(1810)$ (singlet in the quark model). 
The Roper-like  $\Lambda(1600)$ (octet
in the quark model) seems to be missing. This resembles the situation in the $N(\ot^+)$
channel \cite{Engel:2010my}.
The second and third excitations are again dominated by octet interpolators.

\begin{figure}[htb]
\centering
\includegraphics[width=\mylenC,clip]{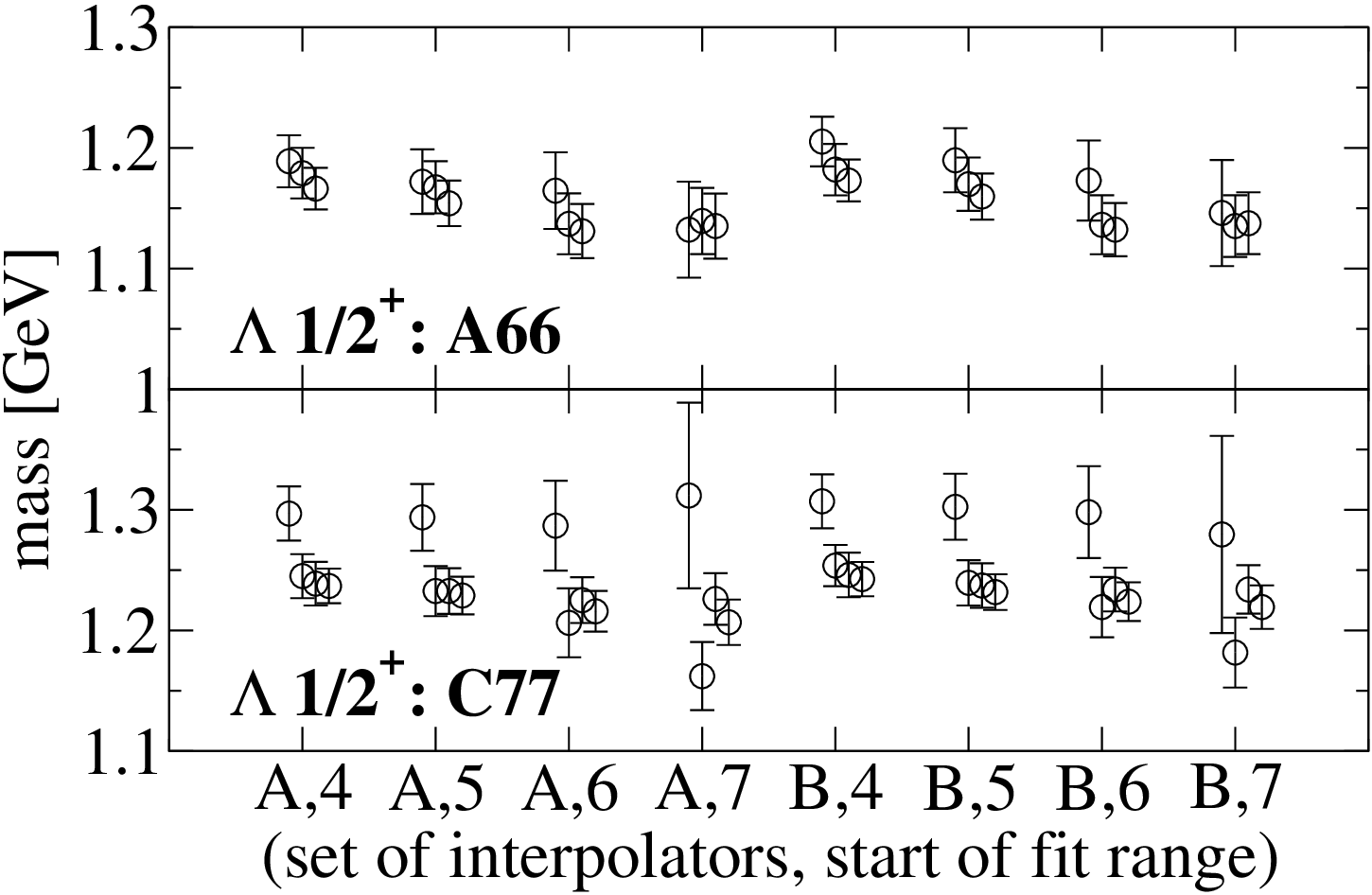}
\caption{Systematic error of the $\Lambda$ $\ot^+$ ground state energy
levels. The levels are shown for different choices of interpolators and
fit ranges, labelled on the horizontal axis. E.g., ``A4'' denotes the set of
interpolator ``A'' and a fit range for the eigenvalues from $t=4a$ to
the onset of noise.  ``A'' denotes interpolators (2,3,10,18,26,27,34,42), ``B''
denotes (3,11,18,27,34). For each set of interpolators and fit range, results
for small to large lattices (spatial size 16, 24 for ensemble A66, and 12,16, 24
for C77, for notation see \cite{Engel:2011aa}) are shown from left to right.
The corresponding  infinite volume limits are the rightmost points.
}
\label{fig:lambda_1half_pospar_vol_syserr}
\end{figure}

\paragraph{Infinite volume extrapolation:}
We performed a volume analysis for several sets of interpolators and various fit
ranges. The results in finite volume and the infinite volume extrapolations for the
ground state for specific interpolators are shown in
Fig.~\ref{fig:lambda_1half_pospar_vol_syserr}. The extrapolation follows the method
of \cite{Durr:2008zz}. A stable choice is the set of interpolators A and
$t_{\text{min}}=5a$.  The corresponding systematic error estimated from different
interpolators and fit ranges is indicated in the summary Fig.~\ref{fig:summary}. Our
final result is $1126(17)(11)$ MeV (statistical and systematic error), which agrees
nicely with  the experimental $\Lambda(1116)$ mass.


\begin{figure}
\noindent\includegraphics[width=\mylenC,clip]{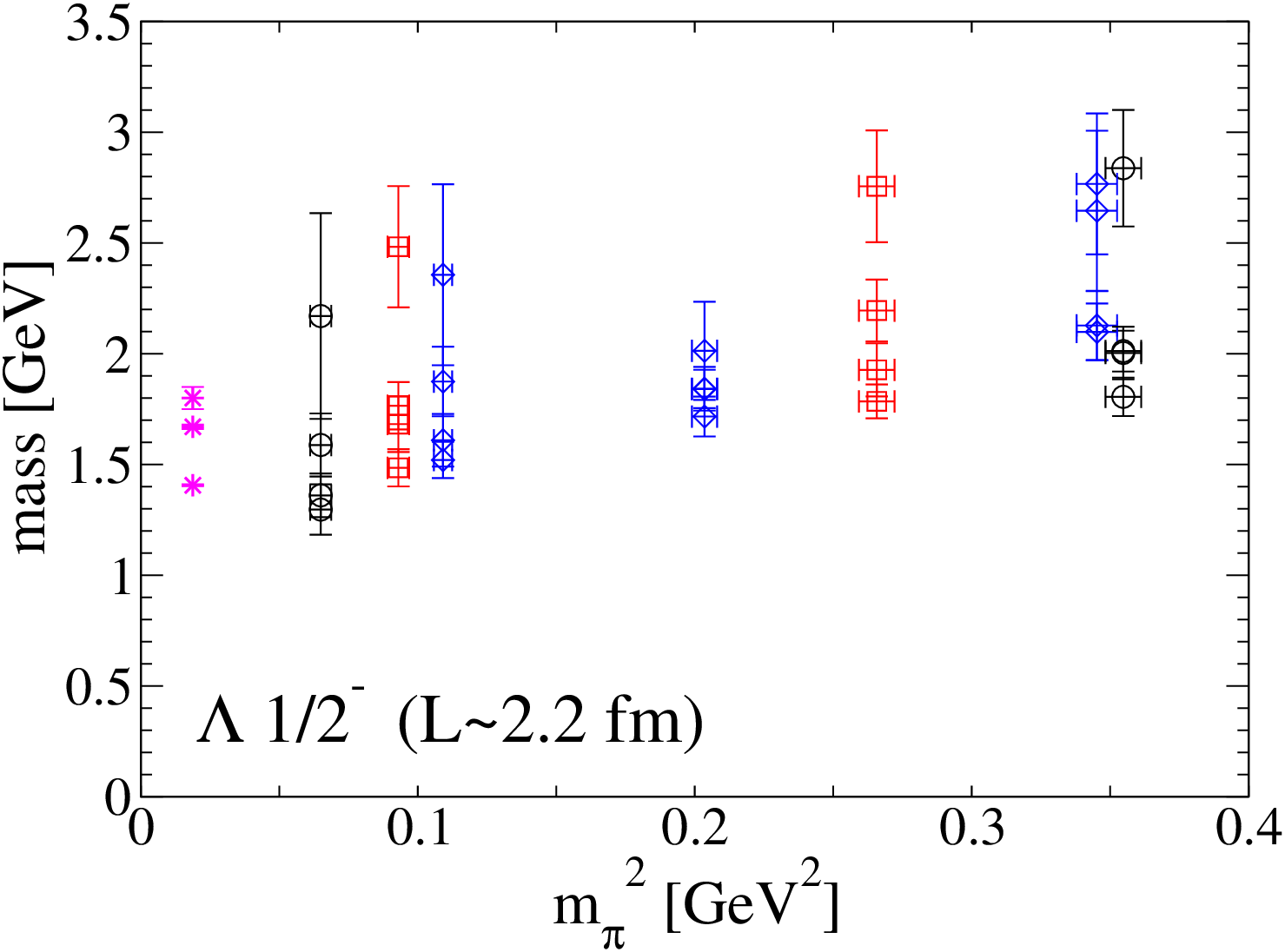}
\caption{Energy levels for $\Lambda$ spin $\ot^-$ in a finite box of linear size $L\approx 2.2$ fm, for legend see caption of Fig.~\ref{fig:lambda_1half_pospar}. Fits are omitted for clarity.}
\label{fig:lambda_1half_negpar}
\end{figure}

\paragraph{$J^P=\ot^{-}$, Finite volume:}
We use different sets of interpolators and fit ranges. We stress that our basis is
large compared to that of other studies with three types of
Dirac structures for both singlet and octet interpolators.  We can extract the lowest
four energy levels, shown in Fig.~\ref{fig:lambda_1half_negpar}, using interpolators
(2,3,10,18,26,27,34,42). We find that the ground state energy level agrees well with
the experimental $\Lambda(1405)$. 
The dependence of the levels on the tuning of the strange quark mass appears to be sizeable, albeit it an accident of our simulation. 
This is one reason of the large systematic error depicted in Fig.~\ref{fig:summary}. 

\begin{figure}
\centering
\noindent\includegraphics[width=\mylenC,clip]{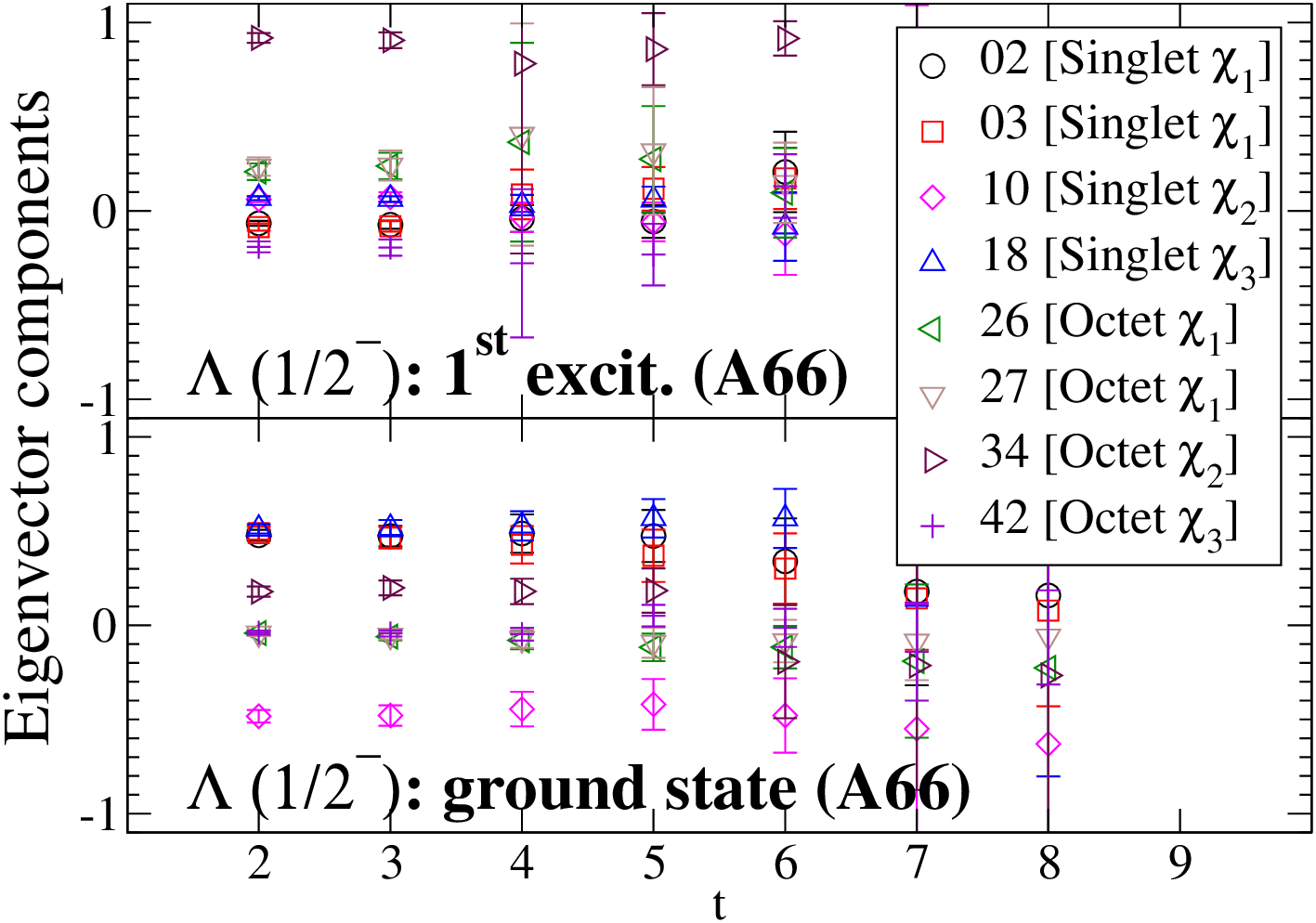}
\caption{Eigenvectors for $\Lambda$ spin $1/2^-$ ground state and first excitation for ensemble A66 ($m_\pi\approx 255$ MeV). 
The ground state is dominated by singlet interpolators of all three Dirac structures. 
The first (and also the second excitation, not shown) is dominated by octet interpolators. 
Note that a considerable mixing of singlet and octet is observed (15-20\% for the ground state). 
}
\label{fig:lambda_1halfneg_vectors}
\end{figure}

The chosen set of interpolators is particularly suitable for an analysis of the
content of the states. The spatial support of the quark fields is
equivalent in all interpolators and hence does not require additional
renormalization when comparing the contribution of different interpolators to the
eigenstate.  In addition, we use several interpolators for each given combination of
flavor and Dirac structures, which allows us to identify a possibly higher number of
states with similar structure. 
Within the basis used, the ground state is dominated by singlet
interpolators of all three Dirac structures. There is, however,  a considerable
mixing with octet interpolators (second Dirac structure) of 15-20\% in ensemble A66,
i.e., at $m_\pi\approx 255$ MeV (see Fig.~\ref{fig:lambda_1halfneg_vectors}). This mixing is expected to increase towards the
physical point, which may complicate the functional dependence of the energy levels on
the pion mass.  The first and second excitation are both dominated by octet
interpolators, by the second and first Dirac structure, respectively.

The first and second excited energy level are both a bit low but compatible with the
experimental $\Lambda(1670)$ and $\Lambda(1800)$. In general in the $J^P=\ot^{-}$ channel one may expect
coupling to $\pi\Sigma$ and $\overline K N$ in $s$-wave. In \cite{Lage:2009zv,Doring:2011ip} the
expected energy levels in finite volumes are discussed
based on a continuum hadron exchange model.
There  (with physical hadron masses), the low-lying scattering state levels in the
$\ot^-$ channel 
are well separated for $m_\pi L\lesssim 3$.
For the pion masses of our study, the non-interacting meson-baryon thresholds
lie close but (except for one point)  above the lowest energy level observed.
E.g., for the lowest pion mass, the values are $m_\Sigma+m_\pi\approx 1.52$ GeV,
$m_N+m_K\approx 1.62$ GeV, above the lowest level. 
This resembles
the situation in the $N(\ot^-)$ channel. Earlier work
argued that the coupling of single baryons to baryon-meson channels may
be too weak to lead to observable effects \cite{Engel:2010my,Bulava:2010yg}.
However, in our case, in $s$-wave scattering, we cannot exclude that one or even
two of the observed three lowest energy levels correspond to scattering states.
Note that the measured ground state energy level is always (except for one point)
below $s$-wave threshold, thus supporting the $\Lambda(1405)$ identification.

It has been conjectured from Chiral Unitary Theory that the lowest
state may have a double-pole \cite{Oller:2000fj,Jido:2003cb} 
and a identification strategy for lattice simulations is suggested in 
\cite{MartinezTorres:2012yi}. This would require asymmetric boxes or 
moving frames.

\paragraph{Infinite volume extrapolation:}
We study the volume dependence of the three lowest states for different sets of
interpolators and various fit ranges.  We emphasize that  the 
stability of the signals of the excitations is comparable to the ones of the ground state. 
The volume dependence of all three low
states appears fairly flat in our simulation, in a few cases showing even negative
finite volume corrections. 
These features are compatible with  significant
contributions of an attractive $s$-wave scattering state.  For interpolators
(2,3,10,18,26,27,34,42) and $t_{\text{min}}=5a$, after infinite volume
extrapolation, we show the result of the final extrapolation of the ground state
energy level to the physical pion mass in Fig.~\ref{fig:summary}.  The final result
for the ground state agrees very well with the experimental $\Lambda(1405)$. Both
the first and the second excitation appear to be a bit low but are compatible with the
experimental $\Lambda(1670)$ and $\Lambda(1800)$ (see
Figs.~\ref{fig:lambda_1half_negpar} and \ref{fig:summary}) and might also possibly be $s$-wave 
$\pi\Sigma$ and $\overline K N$ signals.

\paragraph{$J^P=\frac{3}{2}^{+}$, Finite volume:}
In spin $\frac{3}{2}$ channels, for symmetric quark fields, singlet
interpolators vanish exactly due to Fierz identities. We use different quark
smearing widths and thus circumvent the Fierz identities constructing singlet
interpolators nevertheless.  We derive results for the
lowest three energy levels of the variational analysis of interpolators
(2,9,10,16). Only the ground state can be clearly identified and its
extrapolation agrees with the experimental
$\Lambda(1890)$.  Within the finite basis used, this state is dominated
by octet interpolators.

\begin{figure}
\noindent\includegraphics[width=\mylenC,clip]{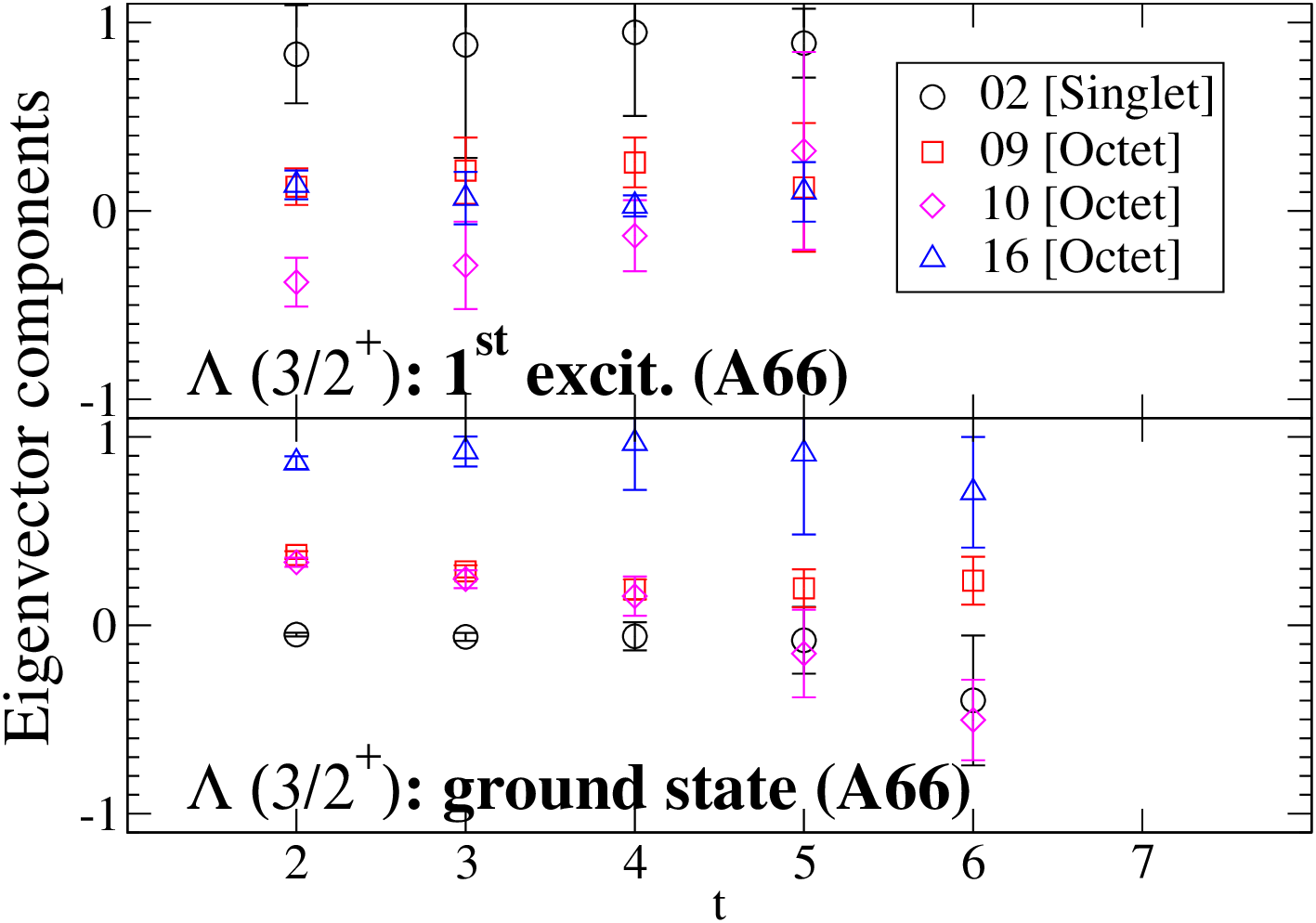}
\caption[Eigenvectors for $\Lambda$ spin $3/2^+$]{
Eigenvectors for $\Lambda$ spin $3/2^+$ ground state and first excitations for ensemble A66 ($m_\pi\approx 255$ MeV). 
We emphasize the domination of singlet interpolators for the first excitation.
Such interpolators are non-vanishing only for broken Fierz identities, which is realized by the use of different quark smearing widths. 
}
\label{fig:lambda_3halfpos_vectors}
\end{figure}

The first excitation is dominated by singlet interpolators with non-negligible
octet contributions at our lightest pion mass (see Fig.~\ref{fig:lambda_3halfpos_vectors}).
We want to emphasize the
importance of singlet interpolators for the low lying states in this channel, 
even though those interpolators are vanishing exactly for symmetric point-like
quark fields. 

\paragraph{Infinite volume extrapolation: }
Within errors we do not observe a clear volume dependence.  The final result
agrees with the experimental $\Lambda(1890)$ mass, but with large uncertainty.

\paragraph{$J^P=\frac{3}{2}^{-}$, Finite volume:}
We choose interpolators (2,7,9,10,15) and find a clear gap between ground state
and excitations. The extrapolation of the ground state energy level lies clearly
above the experimental ground state $\Lambda(1520)$ and is compatible with the
$\Lambda(1690)$. The excitations  extrapolate close to the $\Lambda(2325)$. 

From the eigenvectors we find that the two lowest states are
dominated by octet, the second excitation by singlet interpolators. The quark model
shows for $\Lambda(1520)$ singlet dominance (like for its companion $\Lambda(1405)$). 
We conclude that we might miss a signal for the ground state altogether, or,
alternatively, there is a strong deviation from the leading chiral
behavior towards the physical point.
The latter case is intriguing as it might be related to strong coupling 
to an $s$-wave $\pi\Sigma(1385)$ state, which is discussed, e.g., in 
\cite{Kolomeitsev:2003kt,Sarkar:2004jh}.

\paragraph{Infinite volume extrapolation:}
The volume dependence turns out to be fairly flat, in a few cases even
compatible with negative finite volume corrections.  The final result in the
infinite volume limit at the physical point again misses the  experimental
$\Lambda(1520)$ and agrees with the $\Lambda(1690)$ mass.

\paragraph{Summary:}

We present a comprehensive study of spin $\ot$ and $\frac{3}{2}$ $\Lambda$
baryon ground states and excitations, utilizing a large basis of interpolators
in the  variational analysis including differently smeared quark sources (which allows to
sidestep the Fierz identities), three  Dirac structures,  and singlet and octet forms.  
Fig.~\ref{fig:summary} shows the results for ground states and
excitations for lattices of linear size $L\approx 2.2$ fm (lhs) and
the results for the ground states in the
infinite volume limit (rhs). Systematic errors from the choice of interpolators, the 
fit ranges of the eigenvalues and the tuning of the strange quark mass have been investigated. In both $\ot$ channels 
and in the $\frac{3}{2}^+$ channel we find ground states
extrapolating to the experimental values, in particular we reproduce
$\Lambda(1405)$ and also find two low-lying excitations. 
In our simulation, $\Lambda(1405)$ is dominated by
singlet contributions, but at $m_\pi\approx255$ MeV octet interpolators
contribute roughly 15-20\%, which may increase towards physical pion masses.
The observation of $\Lambda(1405)$ with the employed basis suggests that
this state has a non-negligible singlet 3-quark content. 
The Roper-like (octet) state $\Lambda(1600)$, on the other hand, may not couple to
our 3-quark interpolators.

We analyze the volume dependence and find that only the
ground state of spin $\ot^+$ shows a clear exponential dependence as expected
for bound states.  For all other discussed states, the volume dependence is
either fairly flat or obscured by the statistical error.  For the
$\frac{1}{2}^+$, $\frac{1}{2}^-$ and $\frac{3}{2}^+$ channels the remaining small
deviations at the physical point can be  easily caused by the continuum limit and/or dynamical strange quarks. The discrepancy for the $\frac{3}{2}^-$ ground state might point to
some  effect which is so far not properly accounted for.

\acknowledgments
Special thanks to Leonid Y.~Glozman and Daniel Mohler for valuable criticism.
Discussions with Christof Gattringer, Markus Limmer,
Mario Schr\"ock  and Valentina Verduci are gratefully acknowledged. 
The calculations have been performed on the SGI Altix 4700 of the LRZ Munich and
on local clusters at the University of Graz.  G.P.E.~and A.S.~acknowledge
support by the DFG project SFB/TRR-55.

\bibliographystyle{apsrev4-1}
%

\end{document}